\documentclass[aps,prl,reprint,floatfix
]{revtex4-1} 
\usepackage{colortbl}
\usepackage[T1]{fontenc} 
\usepackage{lmodern}
\usepackage{hyperref}
\usepackage{graphicx}
\usepackage{bm}
\usepackage{amsmath}
\usepackage{amssymb}
\usepackage{dcolumn}
\usepackage{multirow}
\usepackage{color}
\usepackage[dvipsnames]{xcolor}

\newcommand{\heading}[1]{\textbf{#1.---}}

\DeclareMathOperator{\sech}{sech}
\newcommand\picWidthFac{0.75}
\newcommand\picWidthFacTwo{0.825}

\begin{document}
\title{Neutrino Intensity Interferometry:\texorpdfstring{\\}{}Measuring Proto-neutron Star Radii During Core-Collapse Supernovae}

\author{Warren P. \surname{Wright}}
\email{wpwright@ncsu.edu}
\affiliation{Department of Physics,  North Carolina State University, Raleigh, North Carolina 27695, USA}

\author{James P. \surname{Kneller}}
\email{jpknelle@ncsu.edu}
\affiliation{Department of Physics,  North Carolina State University, Raleigh, North Carolina 27695, USA}
\date{\today}

\begin{abstract}
Intensity interferometry is a technique that has been used to measure the size of sources ranging from the quark-gluon plasma formed in heavy ion collisions to the radii of stars. We investigate using the same technique to measure proto-neutron star (PNS) radii with the neutrino signal received from a core-collapse supernovae. Using a full wave-packet analysis, including the neutrino mass for the first time, we derive criteria where the effect can be expected to provide the desired signal, and find that neutrinos from the next Galactic supernova should contain extractable PNS radius information.
\end{abstract}

\maketitle

\heading{Introduction}
This year marks the 30th anniversary of the measurement of the first, and so far only, core-collapse supernova (CCSN) neutrino signal (SN1987A) \cite{1987PhRvL..58.1490H,1987PhRvL..58.1494B,1988PhLB..205..209A,1987EL......3.1315A}. This landmark detection confirmed the general theoretical expectations about CCSN such as the gravitational collapse via neutrino release, the explosion energy and model, and even progenitor mass \cite{2002PhRvD..65f3002L,1987Natur.329..689K,1987PhRvL..58.2722S,2004PhRvD..70d3006C,1987Natur.326..135B,2007fnpa.book.....G}. The neutrino signal was also exploited to derive limits on various neutrino properties such as their mass, charge, magnetic moment, lifetime, and mixing with heavy sterile neutrinos \cite{2007fnpa.book.....G,1987Natur.329...21B,1988PhRvL..60Q1789G,1988PhRvL..61...27B,1989PhRvL..62..505C,1989PhRvL..62..509K,2000NuPhB.590..562D}. The potential to learn even more about neutrinos and supernovae has become a goal of all current and future neutrino detectors: see \cite{2012ARNPS..62...81S} for a recent review. The information in the neutrino signal is encoded in multiple forms: i.e., in the time, energy and flavor. One encoding that is often overlooked is in the spatial distribution of simultaneously detected events. The distribution is nontrivial because the neutrino is a quantum particle and, thus, overlapping wavefunctions can interfere. This phenomenon is known as intensity interferometry.

The tool of intensity interferometry was first proposed by Hanbury-Brown and Twiss (HBT) in 1956 \cite{1956Natur.177...27B} and then used to measure the angular size of Sirius later that year \cite{1956Natur.178.1046H}. This technique has since been applied to a great variety of systems, both bosonic and fermionic (see \cite{1998AcPPB..29.1839B} and references therein). Neutrino intensity interferometry (NII) has been previously considered as a method to distinguish Majorana from Dirac neutrino types \cite{2006PhRvL..96l1802G} and has been applied to investigating interference effects on neutrino signals due to gravitational lensing \cite{2004PhRvD..69f3008C}.

We consider the application of this technique to supernova neutrinos and demonstrate that NII can be used to measure the size of the neutrino source in a CCSN, i.e. the proto-neutron star (PNS). The value of this application was mentioned in \cite{2001NuPhS..91..351P}. Determination of the PNS radius would provide information about the nuclear equation of state describing the structure of the hot, dense stellar material: a major uncertainty in the modeling of core-collapse supernovae (see \cite{2013ApJ...771...51L} and references therein). NII has been previously examined for CCSN neutrinos in \cite{2016arXiv160500344L} where the authors assume that longitudinal spreading of the wave packet (WP) can be ignored. As we demonstrate, the relaxation of that assumption greatly enhances the viability of the technique of NII for supernovae.

The goal of this Letter is to demonstrate the viability of using neutrinos from a Galactic CCSN to do intensity interferometry and, thereby, measure the PNS radius. We start from a more detailed WP formalism than has been previously used in NII and from it derive, for the first time, the conditions necessary for a useful interferometric signal. These conditions relate the neutrino characteristics (energy, mass, and initial WP size) to the source and detector characteristics (source size, source distance and detector event separation). We also discuss detector considerations to determine the feasibility of measuring the two-particle correlations of simultaneous events that are used in intensity interferometry. 

\heading{A Single Neutrino Wave-Packet}
We begin with a standard WP description of the wave function \cite{2004FoPhL..17..103G,2002JHEP...11..017G} for a neutrino with mass $m_{\nu}$. We define $\vec{p}_0$ as the central momentum of the WP, $\sigma_p$ as the momentum uncertainty, $x=(t,\vec{x})$ as the 4-vector spacetime position and $E$ as the neutrino's energy. The Gaussian wave function for a neutrino mass state expressed in natural units is
\begin{align}
	\psi_{\vec{p}_0}\left(x\right)=\int\frac{d^3p}{\left(2\pi\right)^{3/2}}
    \frac{    
    e^{
    	-\frac{\left(\vec{p}-\vec{p}_0\right)^2}{4\sigma_p^2}
    }    
    }{\left(2\pi\sigma_p^2\right)^{3/4}}
    e^{
    	-\bm{i}\left(E\left(\vec{p}\right)t-\vec{p}\cdot\vec{x}\right)
    }.\label{Eqn:BasicWF}
\end{align}
By using an expansion of $E\left(\vec{p}\right)$ around $\vec{p}_0$ to second order, the three dimensional Gaussian integral in Eq. (\ref{Eqn:BasicWF}) can be evaluated \cite{SupplementalMaterial}. The resulting normalized wave function is
\begin{equation}
	\psi_{ij}\equiv\psi_{\vec{p}_{ij}}\left(\vec{x}_{ij},t_{ij}\right)=
		N_{ij} e^{\chi_{ij}},
    \label{Eqn:WF}
\end{equation}
where $\vec{p}_{ij}$ is the central momentum of the neutrino, $\vec{x}_{ij}$ is the displacement corresponding to a neutrino produced at $\vec{r}_i$ and detected at $\vec{d}_j$ and $t_{ij}$ is the time elapsed from when the center of the WP was at $\vec{r}_i$ to when the neutrino was detected at $\vec{d}_j$ (note $t_{ij}$ is not necessarily $\vert\vec{x}_{ij}\vert/v$ with $v$ the velocity corresponding to the momentum $\vec{p}_{ij}$ and mass $m_{\nu}$). 
The quantities $N_{ij}$ and $\chi_{ij}$ are defined to be 
\begin{align}
	N_{ij}=&\left(2\pi\right)^{-3/4}/\left(\sigma_
    		{\perp ij}\sqrt{\sigma_{\parallel ij}}\right)
    	\label{Eqn:WFN}\text{ and}\\
    \chi_{ij}=&
        \bm{i}\left(p_{ij}\cdot x_{ij}\right)
		-\frac{\vec{B}_{ij}^2}{4\sigma_x\sigma_{\perp ij}}
		-\frac{\bm{i}t_{ij}\left(\vec{B}_{ij}\cdot\vec{p}_{ij}\right)^2}
		{8E_i^3\sigma_x^2\sigma_{\perp ij}\sigma_{\parallel ij}}
        \label{Eqn:WFX},
\end{align}
where  
$\sigma_{\perp ij}=\sigma_x+\bm{i}t_{ij}\sigma_p/E_i$, 
$\sigma_{\parallel ij}=\sigma_x+\bm{i}t_{ij}\sigma_p/E_i \gamma_i^2$, 
$2\sigma_x\sigma_p=1$, 
$\vec{B}_{ij}=\vec{x}_{ij}-t_{ij} \vec{p}_{ij}/E_i$, 
$\gamma_i=E_i/m_{\nu}$. Equations (\ref{Eqn:WF}) to (\ref{Eqn:WFX}) may look complicated, but if one assumes $\hat{p}_{ij}=\hat{z}$ then they reduce to the usual description of a sphere-like particle that spreads out into a pancake-like shape as it propagates. In that case, $\sigma_\perp$ is related to the transverse radius of the pancake and $\sigma_\parallel$ to its longitudinal thickness. The growth in the size of the WP for a typical supernova neutrino can be enormous. According to Kersten and Smirnov \cite{2016EPJC...76..339K}, $\sigma_x\sim10^{-11}\text{ cm}$ for supernova neutrinos (similar to \cite{2017arXiv170208338A}). They also calculate that, for a 10-kpc distant SN, a neutrino of energy 15 MeV, and a mass of $\sim0.01\text{ eV}$, the thickness ($\sigma_\parallel$) of the WP at the detector would be $\sim7\text{ m}$ (a similar estimate is obtained from our definition of $\sigma_\parallel$). For the transverse size of the WP at the detector, $\sigma_\perp$, the definitions above result in a size of the order of hundreds of parsecs. This rather significant increase is why the overlap, and thus interference, of neutrino WPs in the detector needs to be considered.

The growth in the size of the WP means that all simultaneous neutrino detections within $\sigma_{\parallel}$ of each other along the line of sight back to the supernova had overlapping WPs. For a meter-scale neutrino detector, and adopting a neutrino flux of $\sim10^{7}\text{m}^{-2}\text{ns}^{-1}$ for the flux at Earth during the neutronization burst for a supernova at $L=10\;{\rm kpc}$ \cite{2016ApJ...817..182W}, we expect $\sim10^{8}$ neutrino WPs in the detector if $\sigma_{\parallel} \sim7\text{ m}$. This expectation of the number of overlapping WPs is significantly larger than previous estimates \cite{2016arXiv160500344L}. The difference is that we have allowed the longitudinal thickness $\sigma_{\parallel}$ of the WP to grow by including the effect of the neutrino mass. As the neutrino mass decreases the growth of the longitudinal dimension of the WP also decreases. 

\heading{Neutrino Intensity Interferometry}
Given that we expect $\sim10^{8}$ neutrinos to be overlapping in the detector, every neutrino detected will have its detection position influenced by interference with many other neutrinos. In order to see this influence we need to detect at least two neutrinos simultaneously, or within a time window during which their WPs overlapped.

Two particle interference is quantified via the two particle wave function, which, for two neutrinos emitted from points $\vec{r}_1$ and $\vec{r}_2$ and detected at $\vec{d}_1$ and $\vec{d}_2$ is 
\begin{align}
	\phi_{\vec{p}_1,\vec{p}_2}\left(\vec{r}_1,\vec{d}_1,\vec{r}_2,\vec{d}_2\right) = 
		\frac{1}{\sqrt{2}}\left(\psi_{11}\psi_{22} \pm\psi_{12}\psi_{21}\right).
    \label{Eqn:TwoParticleWF}
\end{align}
The plus sign applies to a spin singlet, the minus to the spin triplet. In practice the spin triplet dominates because the neutrinos are very relativistic, meaning helicity and handedness are almost identical. The two particle probability density is given by
\begin{align}\begin{aligned}
    \vert\phi\vert^2=&\frac{1}{2}\left(
        \vert\psi_{11}\vert^2\vert\psi_{22}\vert^2
        +\vert\psi_{12}\vert^2\vert\psi_{21}\vert^2
    \right)\\ &\pm\frac{1}{2}\left(
        \psi_{11}^*\psi_{22}^*\psi_{12}\psi_{21}
        +\psi_{12}^*\psi_{21}^*\psi_{11}\psi_{22}
    \right).
    \label{Eqn:TwoParticlePDensity}
\end{aligned}\end{align}
Upon inserting Eq. (\ref{Eqn:WF}) into the interference part (second line) of Eq. (\ref{Eqn:TwoParticlePDensity}), the interference part becomes of the form
\begin{align}
	\pm\frac{N}{2}\left(e^{\chi}+e^{\chi^*}\right)
    =\pm Ne^{\text{Re}\left[\chi\right]}
    	\cos\left(\text{Im}\left[\chi\right]\right)
    \label{Eqn:TwoParticlePDensityCos},
\end{align}
where $\chi=\chi_{11}+\chi_{12}^*+\chi_{21}^*+\chi_{22}$ and $N=N_{11}N_{12}^*N_{21}^*N_{22}=N_{11}^*N_{12}N_{21}N_{22}^*$, which is a good approximation after the WP has had time to spread out considerably. Equation (\ref{Eqn:TwoParticlePDensityCos}) reveals that the two particle density function from Eq. (\ref{Eqn:TwoParticlePDensity}) has the form $\vert\phi\vert^2=f_1\pm f_2\cos\theta$ (where $f_1$ and $f_2$ are functions of the $N_{ij}$'s and $\chi_{ij}$'s from Eq. (\ref{Eqn:WF}) and $\theta=\text{Im}\left[\chi\right]$). If we had used plane waves for the neutrinos (plane waves being a common approach to intensity interferometry) the two particle density function would be $\vert\phi\vert^2=1\pm\cos\theta_\text{HBT}$ \cite{2006PhRvL..96l1802G}. $\theta_\text{HBT}$ is a function of the distance between simultaneously detected events and the dependence of the two-particle correlation upon this distance indicates there are some event separations that are more likely than others. The quantity $\theta=\text{Im}\left[\chi\right]$ that appears in Eq. (\ref{Eqn:TwoParticlePDensityCos}) is also the source of the spatial variation of the separation of simultaneously detected neutrino events. Our goal is to examine how the more general, WP, result for $\theta$ differs from $\theta_\text{HBT}$ in the astrophysical limit as well as any new effects introduced by $f_1$ and $f_2$.

\heading{Two Dimensional Illustration}
In order to appreciate the differences between the WP and plane wave approaches, we adopt the usual assumptions of ballistic momenta $(\hat{p}_{ij}=\hat{x}_{ij}=(\vec{d}_j-\vec{r}_i)/\vert(\vec{d}_j-\vec{r}_i)\vert)$, equal time $\left(t_{ij}=L/v\right)$ and equal energy $\left(E=E_1=E_2\right)$. We also restrict the geometry to a `two-dimensional' case where the emission points $\left(t,x,y,z\right) = \left(0,\pm R,0,0\right)$ and detected locations $\left(L/v,\pm d/2,0,L\right)$ lie in the same spatial plane. In this geometry there are just two path lengths so we define, for the shorter path (which we take to be $r_1 \rightarrow d_1$ and $r_2 \rightarrow d_2$), $\psi_{11}=\psi_{22}\equiv\psi_s=N e^{\chi_s}$ and for the longer path $\psi_{12}=\psi_{21}\equiv\psi_l=N e^{\chi_l}$ (note that $N_s=N_l\equiv N$). We intend to investigate the relaxation of these assumptions (2D point sources, ballistic momenta, equal time and equal energy) in a future work.

\begin{figure}[ht]
	\includegraphics[width=\picWidthFacTwo\linewidth]{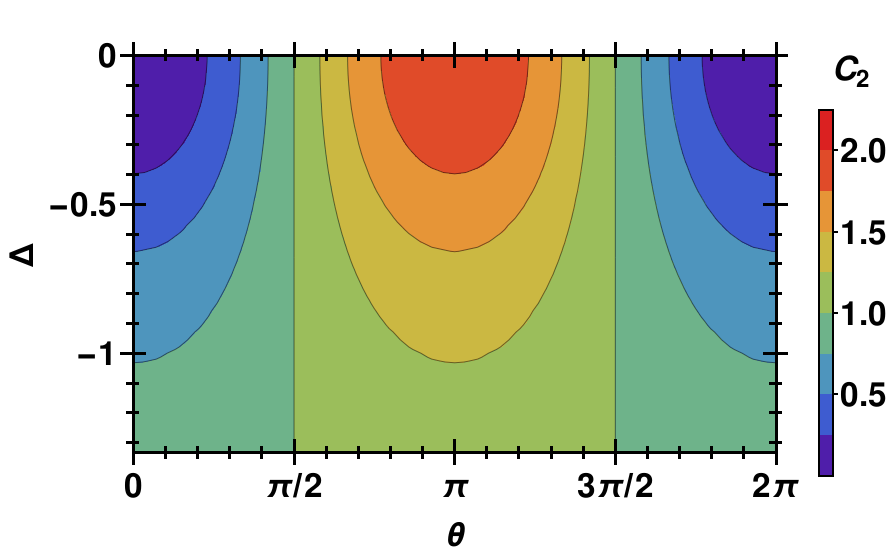}
	\caption{(Color online) Contour plot of $C_2\left(\theta,\Delta\right)$. }
	\label{fig:C2}
\end{figure}

Given a number of neutrino events in a set of time bins, the two point correlation function (2PCF) $C_2(d)$ relates the average number of pairs of events within a bin separated by a distance $d$, $\langle N_2(d)\rangle$, to the average number of events $\langle N_1\rangle$ within the bin:  $\langle N_2(d)\rangle =\langle N_1\rangle^2C_2(d)$. The 2PCF is related to the two particle wavefunction by 
\begin{align}
	C_2=\frac{2\vert\phi\vert^2}{
        \vert\psi_{11}\vert^2\vert\psi_{22}\vert^2
        +\vert\psi_{12}\vert^2\vert\psi_{21}\vert^2
    }.\label{Eqn:C2}
\end{align} 
For the illustrative case described here (see \cite{SupplementalMaterial} for details), the 2PCF takes on a very simple form $C_2=1\pm\cos\theta\sech2\Delta$, where $\theta=2 (\text{Im}\left[\chi_l\right]-\text{Im}\left[\chi_s\right])$ and $\Delta=\text{Re}\left[\chi_l\right]-\text{Re}\left[\chi_s\right]$. For $\Delta=0$ we recover the usual, plane wave, oscillatory correlation pattern \cite{2006PhRvL..96l1802G}, $C_2=1\pm\cos\theta$. For WPs, $\Delta\neq 0$ and as $\vert\Delta\vert$ increases the correlation becomes more and more washed out, i.e. for large $\vert\Delta\vert$, $C_2 \rightarrow 1$. Figure \ref{fig:C2} displays the triplet $C_2\left(\theta,\Delta\right)$ in a contour plot. This figure shows how an increasing $\vert\Delta\vert$ suppresses the interference signal. The origin of the damping factor $\vert\Delta\vert$ is the finite size of the neutrino WP: as the WPs decrease in size, the degree of WP overlap decreases and the correlation disappears. 

The full expressions for $\theta$ and $\Delta$ are quite cumbersome (see \cite{SupplementalMaterial}) but if we assume relativistic neutrinos so that $\gamma^2-1\approx\gamma^2$, and the astrophysical limit, that is $L\gg R\gg d$, we find 
\begin{align}
	\theta&\approx-2\frac{d E R}{L} 
    \left(1+\frac{\gamma ^2 R^2/2}{4 \gamma ^4 E^2 \sigma_x^4+L^2}\right),\label{eqn:theta}\\
    \Delta&\approx-
    \frac{d R^3}{4 L^2 \sigma_x^2}
    \frac{4 E^2 \gamma ^4 \sigma_x^4}{ 4 E^2 \gamma ^4 \sigma_x^4+L^2}.\label{eqn:Delta}
\end{align}

In the massless particle limit ($\gamma\rightarrow\infty$), Eq. (\ref{eqn:theta}) becomes equal to the familiar expression for $\theta_\text{HBT}$ where $\theta_\text{HBT}=-2d E R/L$ \cite{2006PhRvL..96l1802G}. 
To appreciate the scaling of $\theta_\text{HBT}$ with the geometry of the problem and the neutrino energy, we write $\theta_\text{HBT}$ as
\begin{equation}
\theta_\text{HBT}\approx -1\times
	\frac{d}{100\;\text{m}}
	\frac{E}{15\;\text{MeV}}
	\frac{R}{20\;\text{km}}
	\frac{10\;\text{kpc}}{L}.
\end{equation}
Thus we see the probability of detecting simultaneous events will vary on a scale of meters for $E=15\;{\rm MeV}$ neutrinos emitted from a $R=20\;{\rm km}$ PNS at $L=10\;{\rm kpc}$ if they were plane waves. 
The second term (in parentheses) in Eq. (\ref{eqn:theta}) represents the change to $\theta_\text{HBT}$ introduced by the WP formalism. This deviation needs to be small in order for $\theta\approx\theta_\text{HBT}$. If it is large, then $\cos\theta$ oscillates too rapidly with event separation distance $d$ to allow realistic measurement. Similarly, we require that $\Delta$ be small so that the interference occurs and correlation is not overly suppressed. These two requirements give rise to the following inequality, 
\begin{align}
	4\gamma^4E_\nu^2\sigma_x^4+L^2\gg\text{Max}\left\{\frac{\gamma^2R^2}{2}\text{ , }\frac{dR^3\gamma^4E_\nu^2\sigma_x^2}{L^2}\right\}.
\end{align}
This inequality can be rewritten (see \cite{SupplementalMaterial}) in terms of the longitudinal and transverse size of the WP at the detector,
\begin{align}
	\sigma_\parallel^2\gg\frac{d R^3}{4L^2}\quad\text{\;and\;}\quad
	\gamma^2\frac{\sigma_\parallel^2}{\sigma_\perp^2}\gg\frac{R^2}{2L^2}.
    \label{Eqn:Inequalities}
\end{align}
These inequalities are simply conditions relating the WP size to the geometry of the experiment. 

\begin{figure}[ht]
	\includegraphics[width=\picWidthFac\linewidth]{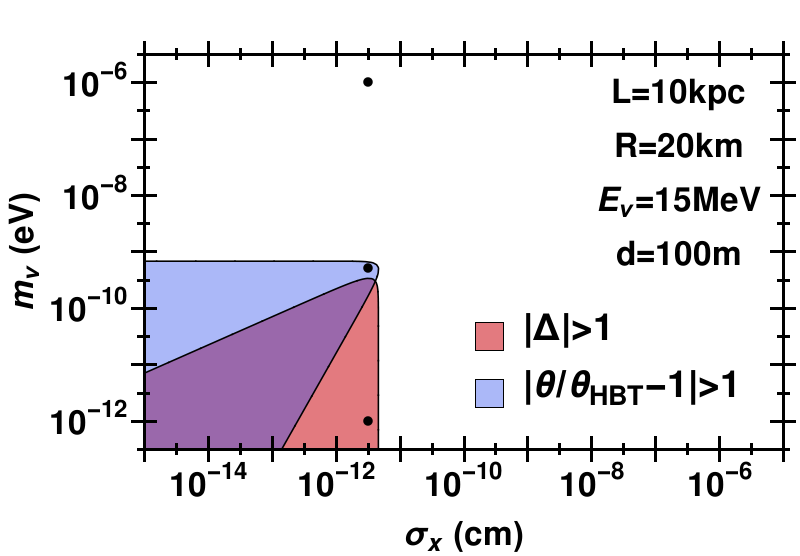}
	\caption{(Color online) Regions in ($m_\nu,\sigma_x$) parameter space where the WP formalism deviates from the plane wave formulation for a typical Galactic CCSN and detector separation. The dots indicate the example values chosen for Fig. \ref{fig:C2vsd}.}
	\label{fig:Regions}
\end{figure}

Let us consider these inequalities for the specific scenario of neutrinos from a Galactic supernova at a distance of $L = 10\;{\rm kpc}$, neutrino energies of $15\;{\rm MeV}$ emitted from opposite sides of a spherical source of radius $R=20\;{\rm km}$, producing two events separated by $d=100\;{\rm m}$. The regions of neutrino mass and WP size where the inequalities are not satisfied are shaded in Fig. \ref{fig:Regions}. 
In this figure, the red region represents the area in parameter space where the correlation damping factor $\Delta$ from Eq. (\ref{eqn:Delta}) suppresses the interference signal and the blue region represents the area where the correlation oscillation frequency from Eq. (\ref{eqn:theta}) deviates from the usual frequency $\theta_\text{HBT}$. From Fig. \ref{fig:Regions}, it is clear that only for very low mass neutrinos which are created with very small initial WP size are the inequalities not satisfied. In the rest of the parameter space the WP treatment gives results which are very similar to the plane wave case. 
 
\begin{figure}[t]
	\includegraphics[width=\picWidthFac\linewidth]{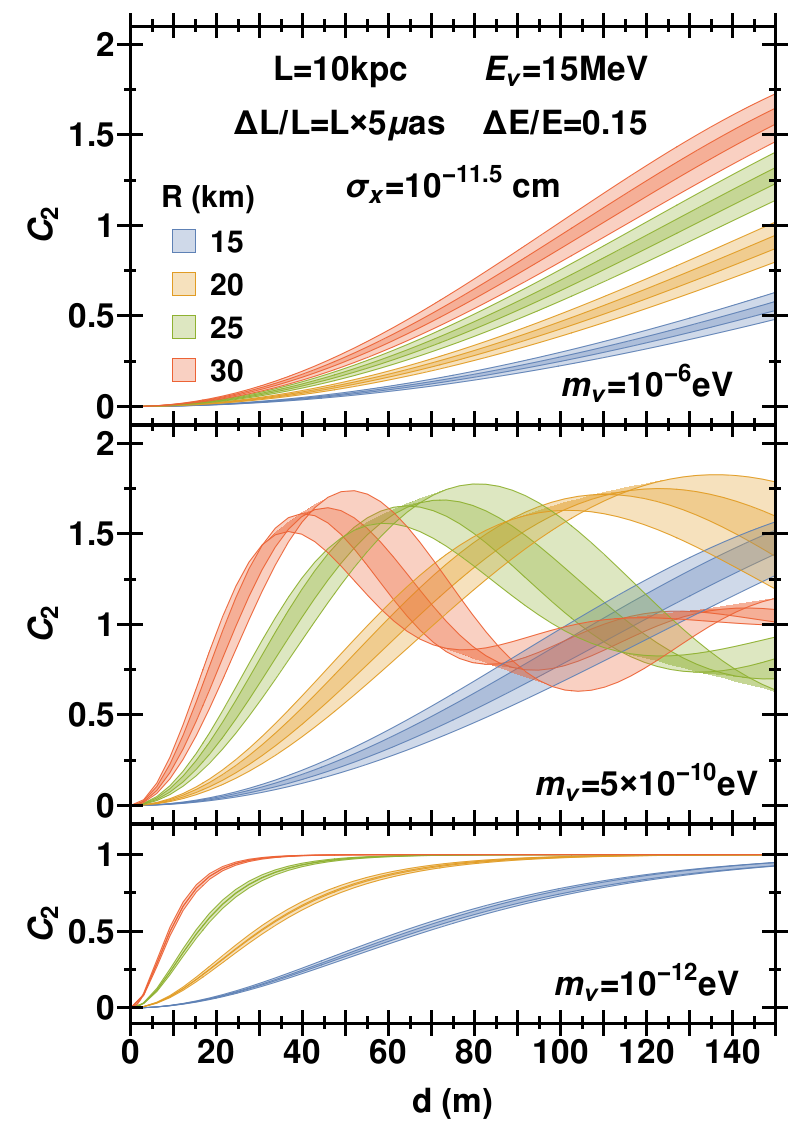}
	\caption{(Color online) A plot of $C_2(d)$ from Eq. (\ref{Eqn:C2}). See text for details.}
	\label{fig:C2vsd}
\end{figure}

Figure \ref{fig:C2vsd} plots triplet $C_2(d)$ from Eq. (\ref{Eqn:C2}) for neutrinos from a Galactic supernova at a distance of $L = 10\;{\rm kpc}$ and neutrino energies of $15\;{\rm MeV}$ emitted from opposite sides of a spherical source with the radii given in the legend. Source distance uncertainty and energy resolution are represented by the inner and outer colored regions respectively. Each of the three subplots corresponds to a different neutrino mass and all subplots are for $\sigma_x=10^{-11.5}\text{ cm}$. These choices for $\left(m_\nu,\sigma_x\right)$ are shown as dots in Fig. \ref{fig:Regions}.

The goal of NII is to determine the source size from a measurement of spatial event correlation. Figure \ref{fig:C2vsd} shows how different size sources give rise to different correlation patterns and, even with realistic distance and energy uncertainties, the differences are distinct. The top subplot represents the most likely case and is simply the usual HBT effect which is unchanged by $\left(m_\nu,\sigma_x\right)$ as long as they satisfy the inequalities in Eq. (\ref{Eqn:Inequalities}).  The middle plot shows the frequency increase due to $\theta>\theta_\text{HBT}$ as well as some damping due to $\vert\Delta\vert$ becoming larger. The bottom plot shows how, as $\vert\Delta\vert$ becomes even larger, the oscillation is washed out as discussed above.  

\heading{Detection Considerations}
The largest difficulty in the practical use of NII is timing resolution. Given an experimental binning time $\tau_\text{bin}$ and a 99\% WP coherence time of $\tau_\text{coh}=3\sigma_\parallel/c$ which is much smaller than $\tau_\text{bin}$, an estimate for the probability of an event pair in an experimental time bin being a HBT-correlated pair is $\sim\tau_\text{coh}/\tau_\text{bin}$ \cite{1998AcPPB..29.1839B}. Multiplying the probability of two events in a time bin and the number of bins, gives the expected number of HBT pairs $N_\text{HBT}$. This number depends upon the neutrino mass and we adopt $m_\nu=0.1$ eV which is well within experimental bounds \cite{Olive:2016xmw}. For a 10 kpc CCSN observed at Hyper-Kamiokande (using data from \cite{2012ARNPS..62...81S}) and $n$ being the number of events in the neutrino detector over a time period of $T$ seconds, we find that $N_\text{HBT}$ is given by (see \cite{SupplementalMaterial})
\begin{align}\begin{aligned}
	N_\text{HBT}\approx
    10^4
    	\left(\frac{n/T}{11000\text{ Hz}}\right)^2
    	\left(\frac{T}{10\text{ s}} \right)
    	\left(\frac{m_\nu}{0.1\text{ eV}}\right)^2
        \\\times
    	\left(\frac{15\text{ MeV}}{E_\nu}\right)^3
    	\left(\frac{10\text{ kpc}}{L}\right)^3
    	\left(\frac{100\text{ fm}}{\sigma_x}\right).
\end{aligned}\end{align}
Using the fiducial values and $n\sim 10^5$ for Hyper-Kamiokande, we see that $\sim 10\%$ of the events in this detector would be HBT correlated. 
Another consideration is detector size and spatial resolution. Given the results shown in Fig. \ref{fig:C2vsd}, it appears we require the detector to be of order several tens of meters in order to distinguish between the 2PCFs of PNSs with different radii. The spatial resolution required of the detector would need to be of the order of meters or better. Both requirements are not unreasonable for current and future detectors \cite{2011arXiv1109.3262A,2016arXiv160301843A}.
As the CCSN moves closer, the necessary size of the detector becomes smaller but also the needed spatial resolution becomes finer. 
Energy resolution is also an important factor but is much more specific to particular detectors. 
Lastly, we briefly comment on the effects of relaxing our assumptions on the detectability of the NII signal. Our expectation is that, given that this technique has been proven to work for photons, relaxing our assumptions could not invalidate the principle of NII, but they could suppress or obfuscate the detected signal (for example, due to a dynamic source and insufficient statistics).

\heading{Conclusion}
We have investigated the possibility of measuring a PNS radius using NII. This investigation used a more complete WP formalism than has been attempted in the past and our simple example has, for the first time, revealed the conditions necessary for a useful interferometric signal. These constraints, shown in Fig. \ref{fig:Regions}, reveal that, for a typical Galactic CCSN, $m_\nu>1\text{ neV}$ or $\sigma_x>100\text{ fm}$ in order for the interferometric signal to be unaltered by WP effects. Detector constraints have also been briefly discussed and detection is plausible in next-generation neutrino detectors. Such a detection would open up a new way of measuring proto-neutron stars and could contribute to the important, but highly nontrivial, determination of the nuclear equation of state. Additionally, should a HBT-type correlation be detected and the other parameters measured to sufficient accuracy, the results derived here could be used to place a lower bound on the neutrino mass.

More work is required to extend this proof of concept to include known neutrino effects that we have ignored. For example, we have assumed the neutrino was produced as a pure mass state which propagated through vacuum. This is clearly not correct and we need to account for the passage through the mantle of the supernova where flavor transformations occur, see, e.g., \cite{2000PhRvD..62c3007D,2006PhRvL..97x1101D,2009PhRvL.103g1101G,2009JPhG...36k3201D,2010PhRvL.104s1102F}. Similarly, we need to account for the flavor structure of the neutrino mass states in the detector and the detector configuration. Such effects will be addressed in a future paper. 

\section{Acknowledgments}
We are grateful to Madappa Prakash for useful discussions and encouragement. 
This work was supported at NC State by DOE grants DE-SC0006417 and DE-FG02-10ER41577.


\bibliographystyle{apsrev4-1}
\bibliography{main}

\end{document}